\documentclass[12pt]{article}
\usepackage[a4paper,margin=0.65in]{geometry}
\usepackage[titletoc,title]{appendix}
\usepackage{graphicx}
\usepackage{amssymb}
\usepackage{lineno}

\usepackage{amsmath}
\usepackage{amsfonts}
\usepackage{amssymb}
\usepackage{algorithm}
\usepackage{algpseudocode}
\usepackage{dsfont}
\usepackage{natbib}
\usepackage{xcolor}
\usepackage{bm}
\usepackage{color}

\renewcommand{\eqref}[1]{Equation~(\ref{#1})}
\newcommand{\Prob}[1]{\mathbb{P}(#1)}
\newcommand{\CondProb}[2]{\mathbb{P}(#1 \mid\nobreak #2)}
\newcommand{\PDF}[1]{p(#1)}
\newcommand{\CondPDF}[2]{p(#1 \mid\nobreak #2)}
\newcommand{\like}[2]{\mathcal{L}(#1 ; #2)}
\newcommand{\Kernel}[2]{K(#1 \mid\nobreak #2)}
\newcommand{\E}[1]{\mathbb{E}\left[#1\right]}
\newcommand{\Econd}[2]{\mathbb{E}\left[#1 \mid\nobreak #2\right]}
\newcommand{\V}[1]{\mathbb{V}\left[#1\right]}
\newcommand{\C}[2]{\mathbb{C}\left[#1,#2\right]}
\newcommand{\bvec}[1]{\mathbf{#1}}

\DeclareMathOperator*{\supp}{supp}
\newcommand{\simProc}[2]{f(#1 \mid\nobreak #2)}
\newcommand{\paramvec}{\boldsymbol{\theta}}
\newcommand{\param}{\theta}
\newcommand{\paramspace}{\boldsymbol{\Theta}}
\newcommand{\discrep}[2]{d( #1, #2)}
\newcommand{\dat}{\mathcal{D}}
\newcommand{\simdat}{\mathcal{D}_s}
\newcommand{\indicator}[2]{\mathds{1}_{#1}(#2)}

\begin{document}

%\begin{frontmatter}



\title{Multilevel rejection sampling for approximate Bayesian computation}

\author{David J. Warne${}^{1\ast}$, Ruth E. Baker${}^{2}$, Matthew J. Simpson${}^{1}$\\
	\\
	\normalsize{${}^{1}$School of Mathematical Sciences, Queensland University of Technology,}\\
	\normalsize{Brisbane, Queensland 4001, Australia}\\
	\normalsize{${}^{2}$Mathematical Institute, University of Oxford,}\\
	\normalsize{Oxford, OX2 6GG, United Kingdom}
	\\
	\normalsize{$^\ast$To whom correspondence should be addressed; E-mail:  david.warne@qut.edu.au.}
}
%\author{David J. Warne\corref{cor}\fnref{QUT}}
%\cortext[cor]{To whom correspondence should be addressed: david.warne@qut.edu.au}
%\author{Ruth E. Baker\fnref{Ox}}
%\author{Matthew J. Simpson\fnref{QUT}}

\date{\today}
\maketitle

%\address[QUT]{School of Mathematical Sciences, Queensland University of Technology, \\
%	Brisbane, Queensland 4001, Australia.}
%\address[Ox]{ Mathematical Institute, University of Oxford, \\
%	Oxford, OX2 6GG, United Kingdom.}
\begin{abstract}
Likelihood-free methods, such as approximate Bayesian computation, are powerful tools for practical inference problems with intractable likelihood functions. Markov chain Monte Carlo and sequential Monte Carlo variants of approximate Bayesian computation can be effective techniques for sampling posterior distributions in an approximate Bayesian computation setting. However, without careful consideration of convergence criteria and selection of proposal kernels, such methods can lead to very biased inference or computationally inefficient sampling. In contrast, rejection sampling for approximate Bayesian computation, despite being computationally intensive, results in independent, identically distributed samples from the approximated posterior. An alternative method is proposed for the acceleration of likelihood-free Bayesian inference that applies multilevel Monte Carlo variance reduction techniques directly to rejection sampling. The resulting method retains the accuracy advantages of rejection sampling while significantly improving the computational efficiency.
\end{abstract}

%\begin{keyword}
%	Bayesian inference \sep approximate Bayesian computation \sep Multilevel Monte Carlo \sep rejection sampling \sep likelihood-free methods.
%	\MSC 62F15 \sep 65C05
%%% keywords here, in the form: keyword \sep keyword
%
%%% PACS codes here, in the form: \PACS code \sep code
%
%%% MSC codes here, in the form: \MSC code \sep code
%%% or \MSC[2008] code \sep code (2000 is the default)
%
%\end{keyword}

%\end{frontmatter}


\section{Introduction}
\label{sec:intro}
Statistical inference is of fundamental importance to all areas of science. Inference enables the testing of theoretical models against observations, and provides a rational means of quantifying uncertainty in existing models. Modern approaches to statistical inference, based on Monte Carlo sampling techniques, provide insight into many complex phenomena~\citep{Beaumont2002,Pooley2015,Ross2017,Stumpf2014,Sunnaker2013,Tavare1997,Thorne2012,Vo2015}.

Suppose we have: a set of observations, $\dat$; a method of determining the likelihood of these observations, $ \like{\paramvec}{\dat}$, under the assumption of some model characterised by parameter vector $\paramvec \in \paramspace$; and a prior probability density, $\PDF{\paramvec}$. The posterior probability density, $\CondPDF{\paramvec}{\dat}$, can be computed using Bayes' Theorem,
\begin{equation}
	\label{eqn:Bayes}
	\CondPDF{\paramvec}{\dat} = \frac{\like{\paramvec}{\dat}\PDF{\paramvec}}{\int_{\paramspace}\like{\paramvec}{\dat}\PDF{\paramvec} \text{d}\paramvec}.
\end{equation}
Explicit expressions for likelihood functions are rarely available~\citep{Tavare1997,Warne2017,Wilkinson2009}; motivating the development of likelihood-free methods, such as approximate Bayesian computation (ABC)~\citep{Stumpf2014,Sunnaker2013}.  ABC methods approximate the likelihood through evaluating the discrepancy between data generated by a simulation of the model of interest and the observations, yielding an approximate posterior,
\begin{equation}
	\label{eqn:approxBayes}
	\CondPDF{\paramvec}{\discrep{\dat}{\simdat} < \epsilon}  \propto \CondProb{\discrep{\dat}{\simdat} < \epsilon}{\paramvec} \PDF{\paramvec}.
\end{equation}
Here, $\mathcal{D}_s \sim \simProc{\mathcal{D}}{\paramvec}$ is data generated by the model simulation process, $\simProc{\mathcal{D}}{\paramvec}$, $d$ is a discrepancy metric, and $\epsilon > 0$ is the acceptance threshold. Due to this approximation, Monte Carlo estimators based on \eqref{eqn:approxBayes} are biased~\citep{Barber2015}. In spite of this bias, however, ABC methods have proven to be very powerful tools for practical inference applications in many scientific areas, including evolutionary biology~\citep{Beaumont2002,Tavare1997,Thorne2012}, ecology~\citep{Stumpf2014}, cell biology~\citep{Ross2017,Johnston2014,Vo2015} and systems biology~\citep{Wilkinson2009}. 

\subsection{Sampling algorithms for ABC}
The most elementary implementation of ABC is ABC rejection sampling~\citep{Pritchard1999,Sunnaker2013}, see Algorithm~\ref{alg:abc-rej}.
This method generates $N$ independent and identically distributed samples $\paramvec^{1},\ldots, \paramvec^{N}$ from the posterior distribution by accepting proposals, $\paramvec^*~\sim~\PDF{\paramvec}$, when the data generated by the model simulation process $\simProc{\dat}{\paramvec^*}$ is within $\epsilon$ of the observed data, $\dat$, under the discrepancy metric $\discrep{\dat}{\cdot}$.
ABC rejection sampling is simple to implement, and samples are independent and identically distributed. Therefore ABC rejection sampling is widely used in many applications~\citep{Browning2017,Grelaud2009,Navascues2017,Ross2017,Vo2015}. However, ABC rejection sampling can be computationally prohibitive in practice~\citep{Barber2015,Fearnhead2012}. This is especially true when the prior density is highly diffuse compared with the target posterior density~\citep{Marin2012}, as most proposals are rejected.
\begin{algorithm}
	\caption{ABC rejection sampler}
	\begin{algorithmic}[1]
		\For{ $i=1,\ldots, N$}
		\Repeat
		\State Sample prior, $\paramvec^* \sim \PDF{\paramvec}$.
		\State Generate data, $\simdat \sim \simProc{\dat}{\paramvec^*}$.
		\Until{$\discrep{\dat}{\simdat} \leq \epsilon$}
		\State Set $\paramvec^{i} \leftarrow \paramvec^{*}$.
		\EndFor
	\end{algorithmic}
	\label{alg:abc-rej}
\end{algorithm}

To improve the efficiency of ABC rejection sampling, one can consider a likelihood-free modification of Markov chain Monte Carlo (MCMC)~\citep{Beaumont2002,Marjoram2003,Tanaka2006} in which a Markov chain is constructed with a stationary distribution identical to the desired posterior. Given the Markov chain is in state $\paramvec^{i}$, a state transition is proposed via a proposal kernel, $\Kernel{\paramvec}{\paramvec^{i}}$.

The Metropolis-Hastings~\citep{Hastings1970,Metroplis1953} state transition probability, $h$, can be modified within an ABC framework to yield
\begin{equation*}
	h = \begin{cases}
		\min \left( \frac{\PDF{\paramvec^{*}}\Kernel{\paramvec^{i}}{\paramvec^{*}}}{\PDF{\paramvec^{i}}\Kernel{\paramvec^{*}}{\paramvec^{i}}},1\right) & \text{if } \discrep{\dat}{\simdat} \leq \epsilon, \\
		0 & \text{otherwise}.
	\end{cases}
\end{equation*}
The stationary distribution of such a Markov chain is the desired approximate posterior~\citep{Marjoram2003}. Algorithm~\ref{alg:mcmc-abc} provides a method for computing $N_T$ iterations of this Markov chain.

While MCMC-ABC sampling can be highly efficient, the samples in the sequence, $\paramvec^1, \ldots, \paramvec^{N_T}$, are not independent. This can be problematic as it is possible for the Markov chain to take long excursions into regions of low posterior probability. This incurs additional bias that is potentially significant~\citep{Sisson2007}. A poor choice of proposal kernel can also have considerable impact upon the efficiency of MCMC-ABC~\citep{Green2015}. The question of how to choose the proposal kernel is non-trivial. Typically proposal kernels are determined heuristically. However, automatic and adaptive schemes are available to assist in obtaining near optimal proposals in some cases~\citep{Cabras2015,Roberts2009}. Another additional complication is that of determining when the Markov Chain has converged; this is a challenging problem to solve in practice~\citep{Roberts2004}.

\begin{algorithm}
	\caption{MCMC-ABC}
	\begin{algorithmic}[1]
		\State{Given initial sample $\paramvec^{1} \sim \CondPDF{\paramvec}{\discrep{\dat}{\simdat} < \epsilon}$.}
		\For{$i = 2, \ldots, N_T$}
		\State Sample transition kernel, $\paramvec^{*} \sim \Kernel{\paramvec}{\paramvec^{i-1}}$.
		\State Generate data, $\mathcal{D}_s \sim \simProc{\mathcal{D}}{\paramvec^*}$.
		\If{$\discrep{\dat}{\simdat} \leq \epsilon$}
		\State Set $h \leftarrow \min \left(\PDF{\paramvec^*}\Kernel{\paramvec^{i-1}}{\paramvec^*}/\PDF{\paramvec^{i-1}}\Kernel{\paramvec^{*}}{\paramvec^{i-1}},1\right)$.
		\State Sample uniform distribution, $u \sim \mathcal{U}(0,1)$.
		\If{$u \leq h$}
		\State Set $\paramvec^{i} \leftarrow \paramvec^{*}$.
		\Else
		\State Set $\paramvec^{i} \leftarrow \paramvec^{i-1}$.
		\EndIf
		\Else
		\State Set $\paramvec^{i} \leftarrow \paramvec^{i-1}$.
		\EndIf
		\EndFor
	\end{algorithmic}
	\label{alg:mcmc-abc}
\end{algorithm}

Sequential Monte Carlo (SMC) sampling was introduced to address these potential inefficiencies~\citep{DelMoral2006} and later extended within an ABC context~\citep{Sisson2007,Drovandi2011,Toni2009}. A set of samples, referred to as particles, is evolved through a sequence of ABC posteriors defined through a sequence of $T$ acceptance thresholds, $\epsilon_1,\ldots,\epsilon_T$~\citep{Sisson2007,Beaumont2009}.
At each step, $t \in [0,T]$, the current ABC posterior, $\CondPDF{\paramvec}{\discrep{\dat}{\simdat} < \epsilon_t}$, is approximated by a discrete distribution constructed from a set of $N_P$ particles $\paramvec_t^{1},\ldots,\paramvec_t^{N_P}$ with importance weights $W_t^{1},\ldots, W_t^{N_P}$ that is, $\Prob{\paramvec = \paramvec_t^i} = W_t^i$ for $i = 1, \ldots, N_P$. The collection is updated to represent the ABC posterior of the next step, $\CondPDF{\paramvec}{\discrep{\dat}{\simdat} < \epsilon_{t+1}}$, through  application of rejection sampling on particles perturbed by a proposal kernel, $\Kernel{\paramvec}{\paramvec^{i-1}}$. If $\CondPDF{\paramvec}{\discrep{\dat}{\simdat} < \epsilon_t}$ is similar to $\CondPDF{\paramvec}{\discrep{\dat}{\simdat} < \epsilon_{t+1}}$, the acceptance rate should be high. The importance weights of the new family of particles are  updated using an optimal backwards kernel~\citep{Sisson2007,Beaumont2009}. Algorithm~\ref{alg:abc-smc} outlines the process.
\begin{algorithm}
	\caption{SMC-ABC}
	\begin{algorithmic}[1]
		\State Initialise $\paramvec_1^{i} \sim \PDF{\paramvec}$ and $W_1^{i} = 1/N_P$, for $i = 1,\ldots, N_P$.
		\For{$t = 2,\ldots, T$}
		\For{$i = 1,\ldots, N_P$}
		\Repeat
		\State Set $\paramvec^* \leftarrow \paramvec_{t-1}^j$ with probability $W_{t-1}^j$.
		\State Sample transition kernel, $\paramvec^{**} \sim \Kernel{\paramvec}{\paramvec^*}$.
		\State Generate data, $\simdat \sim \simProc{\dat}{\paramvec^{**}}$.
		\Until{$\discrep{\dat}{\simdat} \leq \epsilon_t$}
		\State Set $\paramvec_{t}^{i} \leftarrow \paramvec^{**}$.
		\State Set $W_t^{i} \leftarrow \PDF{\paramvec_t^{i}} / \sum_{j=1}^{N_P} W_{t-1}^{j} \Kernel{\paramvec_t^{i}}{\paramvec_{t-1}^{j}}$.
		\EndFor
		\State Normalise weights so that $\sum_{i=1}^{N_P} W_t^{i} = 1$.
		\EndFor
	\end{algorithmic}
	\label{alg:abc-smc}
\end{algorithm}

Through the use of independent weighted particles, SMC-ABC avoids long excursions into the distribution tails that are possible with MCMC-ABC. However, SMC-based methods can be affected by particle degeneracy, and the efficiency of each step is still dependent on the choice of proposal kernel~\citep{Green2015,Filippi2013}. Just as with MCMC-ABC, adaptive schemes are available to guide proposal kernel selection~\citep{Beaumont2009,DelMoral2012}. The choice of the sequence of acceptance thresholds is also important for the efficiency of SMC-ABC. However, there are good solutions to generate these sequences in an adaptive manner~\citep{Drovandi2011,Silk2013}.

\subsection{Multilevel Monte Carlo}
Multilevel Monte Carlo (MLMC) is a recent development that can significantly reduce the computational burden in the estimation of expectations~\citep{Giles2008}. To demonstrate the basic idea of MLMC, consider computing the expectation of a continuous-time stochastic process $X_t$ at time $T$. Let $Z_t^\tau$ denote a discrete-time approximation to $X_t$ with time step $\tau$: the expectations of $X_T$ and $Z_T^\tau$ are related according to
\begin{displaymath}
	\E{X_T} = \E{Z_T^\tau}+\E{X_T - Z_T^\tau}.
\end{displaymath}
That is, an estimate of $\E{Z_T^\tau}$ can be treated as a biased estimate of $\E{X_T}$.
By taking a sequence of time steps $\tau_1 > \cdots > \tau_L$, the indices of which are referred to as \emph{levels}, we can arrive at a telescoping sum,
\begin{equation}
	\label{eqn:mlmc}
	\E{Z_T^{\tau_L}} = \E{Z_T^{\tau_1}} + \sum_{\ell = 2}^L \E{Z_T^{\tau_\ell} - Z_T^{\tau_{\ell-1}}}.
\end{equation}
Computing this form of the expectation returns the same bias as that returned when computing $\E{Z_T^{\tau_L}}$ directly. However, \cite{Giles2008} demonstrates that a Monte Carlo estimator for the telescoping sum can be computed more efficiently than directly estimating $\E{Z_T^{\tau_L}}$ in the context of stochastic differential equations (SDEs).  This efficiency comes from exploiting the fact that the bias correction terms, $\E{Z_T^{\tau_\ell} -Z_T^{\tau_{\ell-1}}}$, measure the expected difference between the estimates on levels $\ell$ and $\ell -1$. Therefore, sample paths of $Z_T^{\tau_{\ell-1}}$ need not be independent of sample paths of $Z_T^{\tau_\ell}$. In the case of SDEs, samples may be generated in pairs driven by the same underlying Brownian motion, that is, the pair is coupled. By the strong convergence properties of numerical schemes for SDEs, \cite{Giles2008} shows that this coupling is sufficient to reduce the variance of the Monte Carlo estimator. This reduction in variance is achieved through optimally trading off statistical error and computational cost across all levels. Through this trade-off, an estimator is obtained with the same accuracy in \emph{mean-square} to standard Monte Carlo, but at a reduced computational cost. This saving of computational cost is achieved since fewer samples of the most accurate discretisation are required.

\subsection{Related work}
Recently, several examples of MLMC applications to Bayesian inference problems have appeared in the literature. One of the biggest challenges in the application of MLMC to inverse problems is the introduction of a strong coupling between levels. That is, the construction of a coupling mechanism that reduces the variances of the bias correction terms enough to enable the MLMC estimator to be computed more efficiently than standard Monte Carlo.  \cite{Dodwall2015} demonstrate a MLMC scheme for MCMC sampling applicable to high-dimensional Bayesian inverse problems with closed-form likelihood expressions. The coupling of \cite{Dodwall2015} is based on correlating Markov chains defined on a hierarchy in parameter space. A similar approach is also employed by \cite{Efendiev2015}.
A multilevel method for ensemble particle filtering is proposed by \cite{Gregory2016} that employs an optimal transport problem to correlate a sequence of particle filters of different levels of accuracy. Due to the computational cost of the transport problem, a local approximation scheme is introduced for multivariate parameters \citep{Gregory2016}. \cite{Beskos2016} look more generally at the case of applying MLMC methods when independent and identically distributed sampling of the distributions on some levels is infeasible, the result is an extension of MLMC in which coupling is replaced with sequential importance sampling, that is, a multilevel variant of SMC (MLSMC).

MLMC has also recently been considered in an ABC context. \cite{Guha2017} extend the work of \cite{Efendiev2015} by replacing the Metropolis-Hasting acceptance probability in a similar way to the MCMC-ABC method~\citep{Marjoram2003}. The MLSMC method~\citep{Beskos2016} is exploited to achieve coupling in an ABC context by \cite{Jasra2017}.

\subsection{Aims and contribution}
ABC samplers based on MCMC and SMC are generally more computationally efficient than ABC rejection sampling~\citep{Marjoram2003,Sisson2007,Toni2009}.  However, there are many advantages to using ABC rejection sampling. Specific advantages are: (i) its simple implementation; (ii) it produces truly independent and identically distributed samples, that is, there is no need to re-weight samples; and (iii) there are no algorithm parameters that affect the computational efficiency. In particular, no proposal kernels need be heuristically defined for ABC rejection sampling.

The aim of this work is to design a new algorithm for ABC inference that retains the aforementioned advantages of ABC rejection sampling, while still being efficient and accurate. The aim is not to develop a method that is always superior to MCMC and SMC methods, but rather provide a method that requires little user-defined configuration, and is still computationally reasonable. To this end, we investigate MLMC methods applied directly to ABC rejection sampling.

Various applications of MLMC to ABC inference have been demonstrated very recently in the literature~\citep{Guha2017,Jasra2017}, with each implementation contributing different ideas for constructing effective control variates in an ABC context. It is not yet clear when one approach will be superior to another. Thus the question of how best to apply MLMC in an ABC context remains an open problem. Since there are no examples of an application of MLMC methods to the most fundamental of ABC samplers, that is rejection sampling, such an application is a significant contribution to this field.

We describe a new algorithm for ABC rejection sampling, based on MLMC. Our new algorithm, called MLMC-ABC, is as general as standard ABC rejection sampling but is more computationally efficient through the use of variance reduction techniques that employ a novel construction for the coupling problem. We describe the algorithm, its implementation, and validate the method using a tractable Bayesian inverse problem. We also compare the performance of the new method against MCMC-ABC and SMC-ABC using a standard benchmark problem from epidemiology.

Our method benefits from the simplicity of ABC rejection sampling. We require only the discrepancy metric and a sequence of acceptance thresholds to be defined for a given inverse problem.
Our approach is also efficient, and achieves comparable or superior performance to MCMC-ABC and SMC-ABC methods, at least for the examples considered in this paper. Therefore, we demonstrate that our algorithm is a promising method that could be extended to design viable alternatives to current state-of-the-art approaches. This work, along with that of \cite{Guha2017} and \cite{Jasra2017}, provides an additional set of computational tools to further enhance the utility of ABC methods in practice.

\section{Methods}
In this section, we demonstrate our application of MLMC ideas to the likelihood-free inference problem given in \eqref{eqn:approxBayes}. The initial aim is to compute an accurate approximation to the joint posterior cumulative distribution function (CDF) of $\paramvec$, using as few data generation steps as possible. We define a data generation step to be a simulation of the model of interest given a proposed parameter vector. While the initial presentation is given in the context of estimating the joint CDF, the method naturally extends to expectations of other functions (see \ref{sec:appdx}). However, the joint CDF is considered first for clarity in the introduction of the MLMC coupling strategy.

We apply MLMC methods to likelihood-free Bayesian inference by reposing the problem of computing the posterior CDF as an expectation calculation. This allows the MLMC telescoping sum idea, as in \eqref{eqn:mlmc}, to be applied. In this context, the levels of the MLMC estimator are parameterised by a sequence of $L$ acceptance thresholds $\epsilon_1,\ldots,\epsilon_L$ with $\epsilon_\ell > \epsilon_{\ell+1}$ for all $\ell \in [1,L)$. The efficiency of MLMC relies upon computing the terms of the telescoping sum with low variance. Variance reduction is achieved through exploiting features of the telescoping sum for CDF approximation, and further computational gains are achieved by using properties of nested ABC rejection samplers.

\subsection{ABC posterior as an expectation}
We first reformulate the Bayesian inference problem (\eqref{eqn:Bayes}) as an expectation. To this end, note that, given a $k$-dimensional parameter space, $\paramspace$, and the random parameter vector, $\paramvec$, if the joint posterior CDF, $F(\bvec{s})~=~\Prob{\param_1 \leq s_1, \ldots, \param_k \leq s_k }$ is differentiable, that is, its probability density function (PDF) exists, then
\begin{align*}
	\Prob{\param_1 \leq s_1, \ldots, \param_k \leq s_k } &= \int_{-\infty}^{s_1}\cdots \int_{-\infty}^{s_k} \CondPDF{\paramvec}{\mathcal{D}} \, \text{d}\param_k\ldots \text{d}\param_1 \\
	&= \int_{A_\bvec{s}}\CondPDF{\paramvec}{\mathcal{D}} \, \text{d}\param_k\ldots \text{d}\param_1,
\end{align*}
where $A_\bvec{s} = \{\paramvec \in \paramspace : \param_1 \leq s_1, \ldots, \param_k \leq s_k\}$.
This can be expressed as an expectation by noting
\begin{align*}
	\int_{A_\bvec{s}}\CondPDF{\paramvec}{\mathcal{D}} \, \text{d}\param_k \ldots \text{d}\param_1 &= \int_{\paramspace}\indicator{A_\bvec{s}}{\paramvec}\CondPDF{\paramvec}{\mathcal{D}} \, \text{d}\param_k \ldots \text{d}\param_1\\
	&= \Econd{\indicator{A_\bvec{s}}{\paramvec}}{\mathcal{D}},
\end{align*}
where $\indicator{A_\bvec{s}}{\paramvec}$ is the indicator function: $\indicator{A_\bvec{s}}{\paramvec} = 1$ whenever $\paramvec \in A_\bvec{s}$ and $\indicator{A_\bvec{s}}{\paramvec} = 0$ otherwise.
Now, consider the ABC approximation given in \eqref{eqn:approxBayes} with discrepancy metric $\discrep{\dat}{\simdat}$ and acceptance threshold $\epsilon$. The ABC posterior CDF, denoted by $F_\epsilon(\bvec{s})$, will be
\begin{align}
	\label{eqn:cdf}
	F_\epsilon(\bvec{s}) &= \int_{A_\bvec{s}} \CondPDF{\paramvec}{\discrep{\dat}{\simdat} < \epsilon} \, \text{d}\param_k\ldots\text{d}\param_1 \notag\\
	&= \Econd{\indicator{A_\bvec{s}}{\paramvec}}{\discrep{\dat}{\simdat} < \epsilon}.
\end{align}
The marginal ABC posterior CDFs, $F_{\epsilon,1}(s), \ldots, F_{\epsilon,k}(s)$, are
\begin{displaymath}
	F_{\epsilon,j}(s) = \lim_{s_{i\neq j} \rightarrow \infty}  F_{\epsilon}(\bvec{s}).
\end{displaymath}

\subsection{Multilevel estimator formulation}
\label{sec:mlmc-est-form}
We now introduce some notation that will simplify further derivations. We define $\paramvec_\epsilon$ to be a random vector distributed according to the ABC posterior CDF, $F_\epsilon(\bvec{s})$, with acceptance threshold, $\epsilon$, as given in \eqref{eqn:cdf}. This provides us with simplification in notation for the ABC posterior PDF, $\PDF{\paramvec_\epsilon} =\CondPDF{\paramvec}{\discrep{\dat}{\simdat} < \epsilon}$, and the conditional expectation,
$\E{\indicator{A_\bvec{s}}{\paramvec_\epsilon}} = \Econd{\indicator{A_\bvec{s}}{\paramvec}}{\discrep{\dat}{\simdat} < \epsilon}$. For any expectation, $P$, we use $\hat{P}^N$ to denote the Monte Carlo estimate of this expectation using $N$ samples.

The standard Monte Carlo integration approach is to generate $N$ samples $\paramvec_\epsilon^{1}, \ldots, \paramvec_\epsilon^{N}$ from the ABC posterior, $\PDF{\paramvec_\epsilon}$, then evaluate the empirical CDF (eCDF),
\begin{equation}
	\label{eqn:ecdf}
	\hat{F}^N_\epsilon(\bvec{s}) = \frac{1}{N} \sum_{i=1}^{N} \indicator{A_\bvec{s}}{\paramvec_\epsilon^{i}},
\end{equation}
for $\bvec{s} \in \mathcal{S}$, where $\mathcal{S}$ is a discretisation of the parameter space~$\paramspace$. For simplicity, we will consider $\mathcal{S}$ to be a $k$-dimensional regular lattice. In general, a regular lattice will not be well suited for high dimensional problems. However, since this is the first time that this MLMC approach has been presented, it is most natural to begin the exposition with a regular lattice, and then discuss other more computationally efficient approaches later. More attention is given to other possibilities in Section~\ref{sec:discuss}.

The eCDF is not, however, the only Monte Carlo approximation to the CDF one may consider. In particular, \cite{Giles2015b} demonstrate the application of MLMC to a univariate CDF approximation.
We now present a multivariate equivalent of the MLMC CDF of \cite{Giles2015b} in the context of ABC posterior CDF estimation. Consider a sequence of $L$ acceptance thresholds, $\{\epsilon_\ell\}_{\ell=1}^{\ell=L}$, that is strictly decreasing, that is, $\epsilon_\ell > \epsilon_{\ell+1}$. In this work, the problem of constructing optimal sequences is not considered, as the focus is the initial development of the method. More discussion around this problem is given in Section~\ref{sec:discuss}. Given such a sequence, $\{\epsilon_\ell\}_{\ell=1}^{\ell=L}$, we can represent the CDF (\eqref{eqn:cdf}) using the telescoping sum
\begin{equation}
	\label{eqn:mlmccdf}
	F_{\epsilon_L}(\bvec{s}) = \E{\indicator{A_\bvec{s}}{\paramvec_{\epsilon_L}}} = \sum_{\ell=1}^{L} Y_\ell(\bvec{s}),
\end{equation}
where
\begin{equation}
	\label{eqn:mlmccdfterms}
	Y_\ell(\bvec{s}) = \begin{cases}
		\E{\indicator{A_\bvec{s}}{\paramvec_{\epsilon_1}}} & \text{if } \ell = 1, \\
		\E{\indicator{A_\bvec{s}}{\paramvec_{\epsilon_\ell}} - \indicator{A_\bvec{s}}{\paramvec_{\epsilon_{\ell -1}}}} & \text{if } \ell > 1.
	\end{cases}
\end{equation}
Using our notation, the MLMC estimator for \eqref{eqn:mlmccdf} and \eqref{eqn:mlmccdfterms} is given by
\begin{eqnarray}
	\hat{F}^{N_1,\ldots,N_L}_{\epsilon_L}(\bvec{s}) = \sum_{\ell = 1}^{L} \hat{Y}^{N_\ell}_\ell(\bvec{s}), \label{eqn:mlmcEst}
\end{eqnarray}
where
\begin{equation}
	\hat{Y}^{N_\ell}_\ell(\bvec{s}) = \begin{cases}
		\frac{1}{N_1} \sum_{i=1}^{N_1} g_\bvec{s}(\paramvec_{\epsilon_1}^{i}) & \text{if } \ell = 1, \label{eqn:bMCest}\\
		\frac{1}{N_\ell} \sum_{i=1}^{N_\ell}\left[ g_\bvec{s}(\paramvec_{\epsilon_\ell}^{i}) - g_\bvec{s}(\paramvec_{\epsilon_{\ell-1}}^{i})\right] & \text{if } \ell > 1,
	\end{cases}
\end{equation}
and $g_\bvec{s}(\paramvec)$ is a Lipschitz continuous approximation to the indicator function; this approximation is computed using a tensor product of cubic polynomials,
\begin{equation*}
	g_\bvec{s}(\paramvec) = \prod_{j=1}^k \xi\left(\frac{s_j - \param_j}{\delta_j}\right),
\end{equation*}
where $\delta_j$ is the lattice spacing in the $j$th dimension, and $\xi(x)$ is a piece-wise continuous polynomial,
\begin{equation*}
	\xi(x) = \begin{cases}
		1 & x \leq -1, \\
		\frac{5}{8}x^3 - \frac{9}{8}x + \frac{1}{2} & -1 < x < 1, \\
		0 & x \geq 1.
	\end{cases}
\end{equation*}
This expression is based on the rigorous treatment of smoothing required in the univariate case given by \cite{Giles2015b}. While other polynomials that satisfy certain conditions are possible~\citep{Giles2015b,Reiss1981}, here we restrict ourselves to this relatively simple form.
Application of the smoothing function improves the quality of the final CDF estimate, just as using smoothing kernels improves the quality of PDF estimators~\citep{Silverman1986}. Such a smoothing is also necessary to avoid convergence issues with MLMC caused by the discontinuity of the indicator function~\citep{Avikainen2009,Giles2015b}.

To compute the $\hat{Y}^{N_1}_1(\bvec{s})$ term (\eqref{eqn:bMCest}), we generate $N_1$ samples $\paramvec_{\epsilon_1}^{1},\ldots, \paramvec_{\epsilon_1}^{N_1}$  from $\PDF{\paramvec_{\epsilon_1}}$; this represents a biased estimate for $F_{\epsilon_L}(\bvec{s})$.
To compensate for this bias, correction terms, $\hat{Y}^{N_\ell}_\ell(\bvec{s})$, are evaluated for $\ell > 2$, each requiring the generation of $N_\ell$ samples $\paramvec_{\epsilon_\ell}^{1}, \ldots, \paramvec_{\epsilon_\ell}^{N_\ell}$  from $\PDF{\paramvec_{\epsilon_\ell}}$ and $N_\ell$ samples  $\paramvec_{\epsilon_{\ell-1}}^{1}, \ldots, \paramvec_{\epsilon_{\ell-1}}^{N_\ell}$  from $\PDF{\paramvec_{\epsilon_{\ell-1}}}$, as given in \eqref{eqn:bMCest}. It is important to note that the samples, $\paramvec_{\epsilon_{\ell-1}}^{1}, \ldots, \paramvec_{\epsilon_{\ell-1}}^{N_\ell}$, used to compute $\hat{Y}^{N_\ell}_\ell(\bvec{s})$ are independent of the samples, $\paramvec_{\epsilon_{\ell-1}}^{1}, \ldots, \paramvec_{\epsilon_{\ell-1}}^{N_{\ell-1}}$, used to compute $\hat{Y}^{N_{\ell-1}}_{\ell-1}(\bvec{s})$. 

\subsection{Variance reduction}
\label{sec:VarReduc}
The goal is to introduce a coupling between levels that controls the variance of the bias correction terms. With an effective coupling, the result is an estimator with lower variance, hence the number of samples required to obtain an accurate estimate is reduced.
Denote $v_\ell$ as the variance of the estimator $\hat{Y}^{N_\ell}_{\ell}(\bvec{s})$. For $\ell \geq 2$ this can be expressed as
\begin{align*}
	v_\ell =& \, \V{g_\bvec{s}(\paramvec_{\epsilon_\ell}) -  g_\bvec{s}(\paramvec_{\epsilon_{\ell-1}})} \\ =& \, \V{g_\bvec{s}(\paramvec_{\epsilon_\ell})} + \V{g_\bvec{s}(\paramvec_{\epsilon_{\ell-1}})} - 2 \cdot \C{g_\bvec{s}(\paramvec_{\epsilon_\ell})}{g_\bvec{s}(\paramvec_{\epsilon_{\ell-1}})},
\end{align*}
where $\V{\cdot}$ and $\C{\cdot}{\cdot}$ denote the variance and covariance, respectively.
Introducing a positive correlation between the random variables $\paramvec_{\epsilon_\ell}$ and $\paramvec_{\epsilon_{\ell-1}}$ will have the desired effect of reducing the variance of $\hat{Y}^{N_\ell}_\ell(\bvec{s})$.

In many applications of MLMC, a positive correlation is introduced through driving samplers at both the $\ell$ and $\ell-1$ level with the same \emph{randomness}. Properties of Brownian motion or Poisson processes are typically used for the estimation of expectations involving SDEs or Markov processes~\citep{Giles2008,Anderson2012,Lester2014}. In the context of ABC methods, however, simulation of the quantity of interest is necessarily based on rejection sampling. The reliance on rejection sampling makes a strong coupling, in the true sense of MLMC, a difficult, if not impossible task. Rather, here we introduce a weaker form of coupling through exploiting the fact that our MLMC estimator is performing the task of computing an estimate of the ABC posterior CDF. We combine this with a property of nested ABC rejection samplers to arrive at an efficient algorithm for computing $\hat{F}_{\epsilon_L}(\bvec{s})$.

We proceed to establish a correlation between levels as follows. Assume we have computed, for some $\ell < L$, the terms $\hat{Y}^{N_1}_1(\bvec{s}),\ldots,\hat{Y}^{N_\ell}_\ell(\bvec{s})$ in \eqref{eqn:mlmcEst}. That is, we have an estimator to the CDF at level $\ell$ by taking the  sum
\begin{displaymath}
	\hat{F}^{N_1,\ldots, N_\ell}_{\epsilon_\ell}(\bvec{s})~=~\sum_{m=1}^\ell \hat{Y}^{N_m}_m(\bvec{s}),
\end{displaymath}
with marginal distributions $\hat{F}^{N_1,\ldots, N_\ell}_{\epsilon_\ell,j}(s_j)$ for $j= 1, \ldots, k$. We can use this to determine a coupling based on matching marginal probabilities when computing $\hat{Y}_{\ell+1}(\bvec{s})$. After generating $N_{\ell+1}$ samples $\paramvec_{\epsilon_{\ell+1}}^1, \ldots, \paramvec_{\epsilon_{\ell+1}}^{N_{\ell+1}}$ from $\PDF{\paramvec_{\epsilon_{\ell+1}}}$, we compute the eCDF, $\hat{F}^{N_{\ell+1}}_{\epsilon_{\ell+1}}(\bvec{s})$, using \eqref{eqn:ecdf} and obtain the marginal distributions, $\hat{F}^{N_{\ell+1}}_{\epsilon_{\ell+1},j}(s_j)$ for $j = 1,\ldots, k$. We can thus generate $N_{\ell+1}$ coupled pairs $\{\paramvec^i_{\epsilon_{\ell+1}},\paramvec^i_{\epsilon_\ell}\}$ by choosing the $\paramvec^i_{\epsilon_\ell}$ with the same marginal probabilities as the empirical probability of $\paramvec^i_{\epsilon_{\ell+1}}$. That is, the $j$th component of $\paramvec^i_{\epsilon_\ell}$ is given by
\begin{displaymath}
	\param^i_{\epsilon_\ell,j} = \hat{G}^{N_1,\ldots,N_\ell}_{\epsilon_\ell}\left(\hat{F}^{N_{\ell+1}}_{\epsilon_{\ell+1},j}(\param^i_{\epsilon_{\ell+1},j})\right),
\end{displaymath}
where $\param^i_{\epsilon_{\ell+1}}$ is the $j$th component of $\paramvec^i_{\epsilon_{\ell+1}}$ and $\hat{G}^{N_1,\ldots,N_\ell}_{\epsilon_\ell , j}(s)$ is the inverse of the $j$th marginal distribution of $\hat{F}^{N_1,\ldots,N_\ell}_{\epsilon_\ell}(\bvec{s})$.  This introduces a positive correlation between the sample pairs, $\{\paramvec_{\epsilon_{\ell+1}}^{i},\paramvec_{\epsilon_{\ell}}^{i}\}$, since an increase in any of the components of $\paramvec_{\epsilon_{\ell+1}}^{i}$ will cause an increase in the same component $\paramvec_{\epsilon_\ell}^{i}$.
This correlation reduces the variance in the bias correction estimator $\hat{Y}^{N_{\ell+1}}_{\epsilon_{\ell+1}}(\bvec{s})$ computed according to \eqref{eqn:bMCest}. We can then update the MLMC CDF to get an improved estimator by using
\begin{displaymath}
	\hat{F}^{N_1,\ldots,N_{\ell+1}}_{\epsilon_{\ell+1}}(\bvec{s}) = \hat{F}^{N_1,\ldots,N_\ell}_{\epsilon_{\ell}}(\bvec{s}) + \hat{Y}^{N_{\ell+1}}_{\epsilon_{\ell+1}}(\bvec{s}),
\end{displaymath}
and apply an adjustment that ensures monotonicity. We continue this process iteratively to obtain $\hat{F}^{N_1,\dots,N_L}_{\epsilon_L}(\bvec{s})$.

It must be noted here that this coupling mechanism introduces an approximation for the general inference problem; therefore some additional bias can be introduced. This is made clear when one considers the process in terms of the copula distributions of $\paramvec_{\epsilon_{\ell +1}}$ and $\paramvec_{\epsilon_{\ell}}$. If these copula distributions are the same, then the coupling is exact and there is no additional bias. The coupling is also exact for the univariate case ($k =1$). Therefore, under the assumption that the correlation structure does not change significantly between levels, then the bias should be low. In practice, this requirement affects the choice of the acceptance threshold sequence, $\epsilon_1, \ldots, \epsilon_L$; we discuss this in more detail in Section~\ref{sec:discuss}. In Section~\ref{sec:tracex}, we demonstrate that for sensible choices of this sequence, the introduced bias is small compared with the bias that is inherent in the ABC approximation.

\subsection{Optimal sample sizes}
\label{sec:optN}
We now require the sample numbers $N_1, \ldots, N_L$ that are the optimal trade-off between accuracy and efficiency. Denote $d_\ell$ as the number of data generation steps required during the computation of $\hat{Y}^{N_\ell}_\ell(\bvec{s})$ and let $c_\ell = d_\ell / N_\ell $ be the average number of data generation steps per accepted ABC posterior sample using acceptance threshold $\epsilon_\ell$.

Given $v_1~=~\V{g_\bvec{s}(\paramvec_{\epsilon_1})}$ and $v_\ell~=~\V{g_\bvec{s}(\paramvec_{\epsilon_\ell}) - g_\bvec{s}(\paramvec_{\epsilon_{\ell-1})}}$, for $\ell~>~1$,  one can construct the optimal $N_\ell$ under the constraint  $\V{\hat{F}^{N_1,\ldots,N_L}_{\epsilon_L}(\bvec{s})} = \mathcal{O}(h^2)$, where $h^2$ is the target variance of the MLMC CDF estimator. As shown by \cite{Giles2008}, using a Lagrange multiplier method, the optimal $N_1,\ldots, N_L$ are given by
\begin{equation}
	\label{eqn:optN}
	N_\ell = \mathcal{O}\left(h^{-2}\right)\sqrt{\frac{v_\ell}{c_\ell}} \sum_{m=1}^L{\sqrt{v_m c_m}}, \quad \ell = 1,\ldots, L.
\end{equation}
In practice, the values for $v_1,\ldots, v_L$ and $c_1,\ldots,c_L$ will not have analytic expressions available; rather, we perform a low accuracy trial simulation with all $N_1 = \cdots = N_L = c $, for some comparatively small constant, $c$, to obtain the relative scaling of variances and data generation requirements.

\subsection{Improving acceptance rates}
\label{sec:accRate}
A MLMC method based on the estimator in~\eqref{eqn:mlmcEst} and the variance reduction strategy given in Section~\ref{sec:VarReduc} would depend on standard ABC rejection sampling (Algorithm~\ref{alg:abc-rej}) for the final bias correction term $\hat{Y}^{N_L}_{\epsilon_L}$. For many ABC applications of interest, the computation of this final term will dominate the computational costs. Therefore, the potential computational gains depend entirely on the size of $N_L$ compared to the number of samples, $N$, required for the equivalent standard Monte Carlo approach (\eqref{eqn:ecdf}). While this approach is often an improvement over rejection sampling, we can achieve further computational gains by exploiting the iterative computation of the bias correction terms.

Let $\supp(f(x))$ denote the support of a function $f(x)$, and note that, for any $\ell \in [2,L]$,  $\supp(\PDF{\paramvec_{\epsilon_\ell}}) \subseteq \supp(\PDF{\paramvec_{\epsilon_{\ell-1}}})$. This follows from the fact that if, for any $\paramvec$, $\CondProb{\discrep{\dat}{\simdat} < \epsilon_\ell}{\paramvec} > 0$, then  $\CondProb{\discrep{\dat}{\simdat} < \epsilon_{\ell-1}}{\paramvec} > 0$ since $\epsilon_\ell < \epsilon_{\ell -1}$. That is, any simulated data generated using parameter values taken from outside $\supp(\PDF{\paramvec_{\epsilon_{\ell-1}}})$ cannot be accepted on level $\ell$ since $\discrep{\dat}{\simdat} > \epsilon_{\ell-1}$ almost surely.  Therefore, we can truncate the prior to the support of $\PDF{\paramvec_{\epsilon_{\ell-1}}}$ when computing $\hat{Y}^{N_\ell}_\ell(\bvec{s})$, thus increasing the acceptance rate of level $\ell$ samples. In practice, we need to approximate the support regions through sampling. For simplicity, in this work we restrict sampling of the prior at level $\ell$ to within the bounding box that contains all the samples generated at level $\ell-1$. However, more sophisticated approaches could be considered, and may result in further computational improvements.

\subsection{The algorithm}
\label{sec:mlmc-algs}
We now have all the components to construct our MLMC-ABC algorithm. We compute the MLMC CDF estimator (\eqref{eqn:mlmcEst}) using the coupling technique for the bias correction terms (Section~\ref{sec:VarReduc}) and prior truncation for improved acceptance rates (Section~\ref{sec:accRate}).

Optimal $N_1,\ldots, N_L$ are estimated as per \eqref{eqn:optN} and Section~\ref{sec:optN}. Once $N_1,\ldots, N_L$ have been estimated, computation of the MLMC-ABC posterior CDF $\hat{F}_{\epsilon_L}(\bvec{s})$ proceeds according to Algorithm~\ref{alg:mlmc-abc}.

\begin{algorithm}
	\caption{MLMC-ABC}
	\begin{algorithmic}[1]
		\State Initialise $\epsilon_1,\ldots, \epsilon_L$, $N_1,\dots, N_L$ and prior $\PDF{\paramvec}$.
		\State Set $\PDF{\paramvec_{\epsilon_0}} \leftarrow \PDF{\paramvec}$.
		\For{$\ell = 1,\ldots, L$}
		\For{$i = 1, \ldots, N_\ell$}
		\Repeat
		\State Sample $\paramvec^{*} \sim \PDF{\paramvec}$ restricted to $\supp(\PDF{\paramvec_{\epsilon_{\ell-1}}})$.		
		\State Generate data, $\simdat \sim \simProc{\dat}{\paramvec^*}$.
		\Until{$\discrep{\dat}{\simdat} \leq \epsilon_\ell$.}
		\State Set $\paramvec_{\epsilon_\ell}^{i} \leftarrow \paramvec^{*}$.
		\EndFor
		\For{$\bvec{s} \in \mathcal{S}$}
		\State Set $\hat{F}_{\epsilon_\ell}^{N_\ell}(\bvec{s}) \leftarrow \sum_{i=1}^{N_\ell} g_\bvec{s}(\paramvec_{\epsilon_\ell}^{i})/N_\ell$.
		\EndFor
		\If{$\ell > 1$}
		\For{$i = 1, \ldots, N_\ell$}
		\For{$j = 1,\ldots, k$}
		\State Set $\param_{\epsilon_{\ell-1},j}^{i} \leftarrow \hat{G}^{N_1,\dots,N_{\ell-1}}_{\epsilon_{\ell-1},j}\left(\hat{F}_{\epsilon_\ell,j}^{N_\ell}\left(\param_{\epsilon_\ell,j}^{i}\right)\right)$.
		\EndFor
		\EndFor
		\For{$\bvec{s} \in \mathcal{S}$}
		\State Set $\hat{Y}^{N_\ell}_{\epsilon_\ell}(\bvec{s}) \leftarrow \sum_{i=1}^{N_\ell} \left[g_\bvec{s}(\paramvec_{\epsilon_\ell}^{i}) - g_\bvec{s}(\paramvec_{\epsilon_{\ell-1}}^{i})\right]/N_\ell$.
		\State Set $\hat{F}^{N_1,\ldots,N_\ell}_{\epsilon_\ell}(\bvec{s}) \leftarrow \hat{F}^{N_1,\ldots,N_{\ell-1}}_{\epsilon_{\ell-1}}(\bvec{s})+ \hat{Y}^{N_\ell}_{\epsilon_\ell}(\bvec{s})$.
		\EndFor
		\EndIf
		\EndFor
	\end{algorithmic}
	\label{alg:mlmc-abc}
\end{algorithm}

The computational complexity of MLMC-ABC (Algorithm~\ref{alg:mlmc-abc}) is roughly $\mathcal{O}(c_S + N_L(c_L + c_{G}))$ where $c_L$ is the expected cost of generating a single sample of the ABC posterior with threshold $\epsilon_L$, $c_S$ is the cost of updating the CDF estimate and $c_{G}$ is the cost of the coupling which involves the marginal CDF inverses. From Algorithm~\ref{alg:mlmc-abc}, we have $c_S = \mathcal{O}(N_1|\mathcal{S}|)$ and from \cite{Barber2015} we have $c_L = \mathcal{O}(\epsilon^{-n})$ where $n$ is the dimensionality of $\dat$. The marginal inverse operations require only two steps:
\begin{enumerate}
	\item find $\bvec{s} \in \mathcal{S}$ such that, $s_j \leq \param_{\epsilon_\ell,j}^i < s_j + \delta_j$. Such an operation is, at most, $\mathcal{O}(\log_k |\mathcal{S}|)$;
	\item invert the interpolating cubic spline, which can be done in $\mathcal{O}(1)$.
\end{enumerate}
It follows that $c_G = \mathcal{O}(\log_k |\mathcal{S}|)$.
For any practical application, $\epsilon_L$ will need to be sufficiently small, that is, $(N_1|\mathcal{S}|/N_L - \log_k |\mathcal{S}|) \ll c_L$, in order for the cost of generating posterior samples at level $L$ dominates the cost of the marginal CDF inverse operations and the lattice updating. Computational gains over ABC rejection sampling are achieved through decreasing $N_L$ and $c_L$, via variance reduction and prior truncation.

Our primary focus in Algorithm~\ref{alg:mlmc-abc} is on using MLMC to estimate the posterior CDF. The coupling mechanism is more readily communicated in this case. However, the MLMC-ABC method is more general and can be used to estimate $\E{U(\paramvec_{\epsilon_\ell})}$ where $U(\paramvec_{\epsilon_\ell})$ is any Lipschitz continuous function. In this more general case, only the marginal CDFs need be accumulated to facilitate the coupling mechanism. For more details see \ref{sec:appdx}.

\section{Results}
In this section, we provide numerical results to demonstrate the validity, accuracy and performance of our MLMC-ABC method using some common models from epidemiology. In the first instance, we consider a tractable compartmental model, the stochastic SIS (Susceptible-Infected-Susceptible) model~\citep{Weiss1971}, to confirm the convergence and accuracy of MLMC-ABC. We then consider the Tuberculosis transmission model introduced by~\cite{Tanaka2006} as a benchmark to compare our method with MCMC-ABC~(Algorithm~\ref{alg:mcmc-abc}) and SMC-ABC (Algorithm~\ref{alg:abc-smc}).  While we have chosen to perform our evaluation using an epidemiological model due to the particular prevalence of ABC methods in the life sciences, the techniques outlined in this manuscript are completely general and  applicable to many areas of science.

\subsection{A tractable example}
\label{sec:tracex}
The SIS model is a common model from epidemiology that describes the spread of a disease or infection for which no significant immunity is obtained after recovery; the common cold, for example. The model is given by
\begin{eqnarray*}
	S+I \overset{\beta}{\rightarrow} 2I, \\
	I \overset{\gamma}{\rightarrow} S,
\end{eqnarray*}
with parameters $\paramvec = \{\beta,\gamma\}$ and hazard functions for infection and recovery given by
\begin{displaymath}
	h_I(S,I) =\beta S I \text{ and } h_R(S,I) = \gamma I,
\end{displaymath} respectively. This process defines a discrete-state continuous-time Markov process with a forward transitional density function that is computationally feasible to evaluate exactly for small populations $N_{pop} = S + I$. That is, for $t > s$, the probability of $S(t) = x$ given $S(s) = y$, with density denoted by
$\CondPDF{x,t}{y,s; \beta, \gamma}$, has a solution obtained through the $x$th element of the vector
\begin{equation}
	P(y,\beta,\gamma) = \exp(Q(\beta,\gamma)(t-s))\bvec{y},
	\label{eqn:sisbke}
\end{equation}
where the $y$th element of column vector, $\bvec{y}$, is one and all other elements zero, $\exp(\cdot)$ denotes the matrix exponential and $Q(\beta,\gamma)$ is the infinitesimal generator matrix of the Markov process; for the SIS model, $Q(\beta,\gamma)$ is a tri-diagonal matrix dependent only on the parameters of the model.

Let $S_{obs}(t)$ be an observation at time $t$ of the number of susceptible individuals in the population. We generate observations using a single realisation of the SIS model with parameters $\beta = 0.003$ and $\gamma = 0.1$, population size $N_{pop}=101$, and initial conditions $S(0) = 100$ and $I(0) = 1$;  Observations are taken at times $t_1 = 4,\, t_2 = 8,\ldots,\, t_{10} = 40$. Using the analytic solution to the SIS transitional density (using \eqref{eqn:sisbke}), we arrive at the likelihood function
\begin{equation}
	\like{\beta,\gamma}{\dat} = \prod_{i=1}^{10}\CondPDF{S_{obs}(t_i),t_i}{S_{obs}(t_{i-1}),t_{i-1}; \beta, \gamma},
	\label{eqn:sislike}
\end{equation}
where $S_{obs}(t_0) = s_0$ almost surely. Hence, we can obtain an exact solution to the SIS posterior $\CondPDF{\beta,\gamma}{\dat}$ given the priors $\beta \sim \mathcal{U}(0,0.06)$ and $\gamma \sim \mathcal{U}(0,2)$. Given this exact posterior, quadrature can be applied to compute the posterior CDF, given by
\begin{equation*}
	F(s_1,s_2) = \int_{-\infty}^{s_1}\int_{-\infty}^{s_2} \CondPDF{\beta,\gamma}{\dat}\, \text{d}\beta \text{d} \gamma.
\end{equation*}

For the ABC approximation we use a discrepancy metric based on the sum of squared errors,
\begin{equation*}
	\discrep{\dat}{\simdat} = \left[\sum_{i=1}^{10} \left(S_{obs}(t_i) - S^*(t_i)\right)^2\right]^{1/2},
\end{equation*}
where $S^*(t)$ is a realisation of the model generated using the Gillespie algorithm~\citep{Gillespie1977} for a given set of parameters, and $\simdat = [S^*(t_1),S^*(t_2),\ldots,S^*(t_{10})]$. An appropriate acceptance threshold sequence for this metric is $\epsilon_1, \ldots, \epsilon_L$, with $\epsilon_\ell = \epsilon_1 m^{1-\ell}$, $m = 2$ and $\epsilon_1 = 75$~\citep{Toni2009}.

We can use the exact likelihood (\eqref{eqn:sislike}) to evaluate the convergence properties of our MLMC-ABC method. Based on the analysis of ABC rejection sampling by~\cite{Barber2015}, the rate of decay of the root mean-squared error (RMSE) as the computational cost is increased is slower for ABC than standard Monte Carlo. We find experimentally for ABC rejection sampling, that the decay of RMSE of the SIS model is approximately $\mathcal{O}(C^{-1/4})$, where $C$ is the expected computational cost of generating $N$ ABC posterior samples. The $95\%$ confidence interval of this experimental rate is $[-0.27, -0.23]$, which is consistent with the expected theoretical result of $-0.25$~\citep{Barber2015}. This is slower than the expected $\mathcal{O}(C^{-1/2})$ decay typically achieved with standard Monte Carlo. We compare the ABC rejection sampler decay rate with that achieved by our MLMC-ABC methods.

\begin{figure}[h]
	\centering	
	\includegraphics[width=0.65\columnwidth]{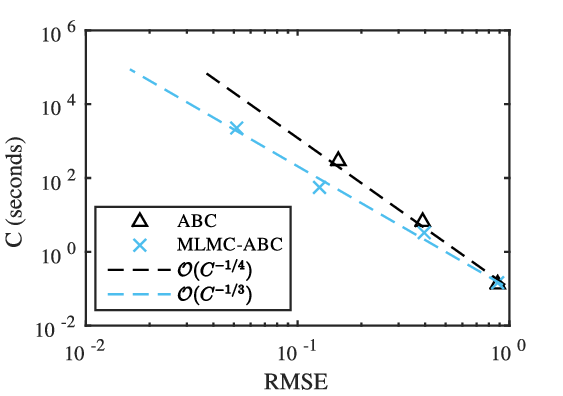}
	\caption{RMSE convergence for MLMC-ABC compared with ABC rejection sampling. The RMSE is computed using the exact solution to the posterior density of the SIS model.}
	\label{fig:conv}
\end{figure}

We use the $L_\infty$ norm for the RMSE, that is,
\begin{displaymath}
	\text{RMSE} = \sqrt{\E{\left\|F - \hat{F}_{\epsilon_L}\right\|_\infty^2}},
\end{displaymath}
where $F$ is the exact posterior CDF, evaluated directly from the likelihood (\eqref{eqn:sislike}), and $\hat{F}_{\epsilon_L}$ is the Monte Carlo estimate of the CDF. A regular lattice that consists of $300 \times 300$ points is used for this estimate. The computation required for Gillespie simulations completely dominates the added computation of updating the lattice.
The RMSE is computed using $20$ independent MLMC-ABC CDF estimations for $L = 1,\ldots, 3$. The sequence of sample numbers, $N_1,\ldots,N_L$, is computed using \eqref{eqn:optN} with target variance of $\mathcal{O}(\epsilon^2)$ and $100$ trial samples. Figure~\ref{fig:conv} demonstrates the improved convergence rate over the ABC rejection sampler convergence rate. Based on the least-squares fit, the RMSE decay is approximately $\mathcal{O}(C^{-1/3})$. The $95\%$ confidence interval for the rate is $[-0.26, -0.34]$ which is consistent with theory from~\cite{Giles2015b} for the univariate situation. For smaller target RMSE, this results in an order of magnitude reduction in the computational cost.

\begin{figure}[h]
	\centering	
	\includegraphics[width=0.65\columnwidth]{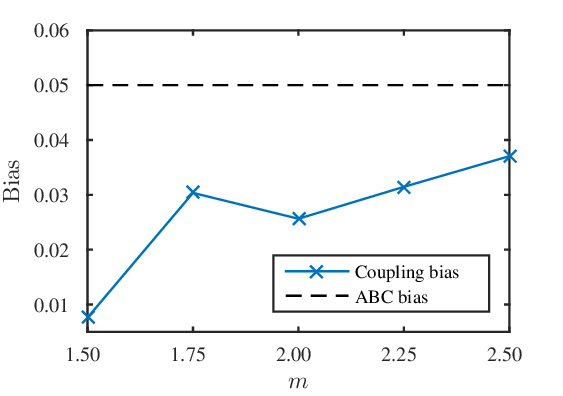}
	\caption{Convergence of coupling bias as a function of $m$.}
	\label{fig:bias}
\end{figure}

We also consider the effect of the additional bias introduced through the coupling mechanism as presented in Section~\ref{sec:VarReduc}. We set the sample numbers at all levels to be $10^4$ to ensure the Monte Carlo error is negligible and compare the bias  for different values of acceptance threshold scaling factor $m$. The bias is computed according to the $L_\infty$ norm, that is,
\begin{displaymath}
	\text{Bias} =\E{\left\|\hat{F}_{\epsilon_L}^c - \hat{F}_{\epsilon_L}^u\right\|_\infty},
\end{displaymath}
where $\hat{F}_{\epsilon_L}^c$ denotes the estimator computed according to Algorithm~\ref{alg:mlmc-abc} and $\hat{F}_{\epsilon_L}^u$ denotes the estimator computed without any coupling. That is, $\hat{F}_{\epsilon_L}^u$ is computed using standard Monte Carlo to evaluate each term in the MLMC telescoping sum (\eqref{eqn:mlmccdf}). Note that, computationally, $\hat{F}_{\epsilon_L}^u$ will always be inferior to a standard Monte Carlo estimate. Figure~\ref{fig:bias} shows that, not only does the bias decay as $m$ decreases, but also the additional bias is well within the order of bias expected from the ABC approximation.
This result, along with Figure~\ref{fig:conv}, demonstrates that the reduction in estimator variance can dominate the increase in bias. Thus, compared with standard Monte Carlo, a significantly lower RMSE for the same computational effort is achieved with MLMC using this coupling strategy.

\subsection{Performance evaluation}
We now evaluate the performance of MLMC-ABC using the model developed by \cite{Tanaka2006} in the study of tuberculosis transmission rate parameters using DNA fingerprint data~\citep{Small1994}. This model has been selected due to the availability of published comparative performance evaluations of MCMC-ABC and SMC-ABC~\citep{Tanaka2006,Sisson2007}.

The model proposed by \cite{Tanaka2006} describes the occurrence of tuberculosis infections over time and the mutation of the bacterium responsible, \emph{Myobacterium tuberculosis}. The number of infections, $I$, at time $t$ is
\begin{displaymath}
	I_t = \sum_{i=1}^{G_t} X_{i,t},
\end{displaymath}
where $G_t$ is the number of distinct genotypes  and $X_{i,t}$ is the number of infections caused by the $i$th genotype at time $t$. For each genotype, new infections occur with rate $\alpha$, infections terminate with rate $\delta$, and mutation occurs with rate $\mu$; causing an increase in the number of genotypes. This process, as with the SIS model, can be described by a discrete-state continuous-time Markov process. In this case, however, the likelihood is intractable, but the model can still be simulated using the Gillespie algorithm~\citep{Gillespie1977}. After a realisation of the model is completed, either by extinction or when a maximum infection count is reached, a sub-sample of $473$ cases is collected and compared against the IS$6110$ DNA fingerprint data of tuberculosis bacteria samples~\citep{Small1994}. The dataset consists of $326$ distinct genotypes; the infection cases are  clustered according to the genotype responsible for the infection. The collection of clusters can be summarised succinctly as $
30^1 \, 23^1 \, 15^1 \, 10^1 \, 8^1 \, 5^2 \, 4^4 \,  3^{13} \, 2^{20} \, 1^{282},$
where $n^j$ denotes there are $j$ clusters of size $n$. The discrepancy metric used is
\begin{equation}
	\label{eqn:tb_disc}
	\discrep{\dat}{\simdat} = \frac{1}{n}\left|g(\dat) - g(\simdat) \right| + \left| H(\dat) - H(\simdat)\right|,
\end{equation}
where $n$ is the size of the population sub-sample (e.g., $n = 473$), $g(\dat)$ denotes the number of distinct genotypes in the dataset (e.g., $g(\dat) = 326$) and the genetic diversity is \[H(\dat) = 1 - \frac{1}{n^2}\sum_{i=1}^{g(\dat)} n_i(\dat)^2,\]
where $n_i(\dat)$ is the cluster size of the $i$th genotype in the dataset.

We perform likelihood-free inference on the tuberculosis model for the parameters $\paramvec = \left\{\alpha,\delta,\mu\right\}$ with the goal of evaluating the efficiencies of MLMC-ABC, MCMC-ABC and SMC-ABC. We use a target posterior distribution of $\CondPDF{\paramvec}{\discrep{\dat}{\simdat} < \epsilon}$ with $\discrep{\dat}{\simdat}$ as defined in \eqref{eqn:tb_disc} and $\epsilon = 0.0025$. The acceptance threshold sequence, $\epsilon_1,\ldots, \epsilon_{10}$, used for both SMC-ABC and MLMC-ABC is $\epsilon_i = \epsilon_{10} + (\epsilon_{i-1} -\epsilon_{10})/2$ with $\epsilon_1 = 1$ and $\epsilon_{10} = 0.0025$. The improper prior is given by  $\alpha \sim \mathcal{U}(0,5)$, $\delta \sim \mathcal{U}(0,\alpha)$ and $\mu \sim \mathcal{N}(0.198,0.06735^2)$~\citep{Sisson2007,Tanaka2006}. For the MCMC-ABC and SMC-ABC algorithms we apply a typical Gaussian proposal kernel,
\begin{displaymath}
	\Kernel{\paramvec^{(i)}}{\paramvec^{(i-1)}} = \mathcal{N}\left(\paramvec^{(i-1)},\Sigma\right),
\end{displaymath}
with covariance matrix
\begin{eqnarray}
	\label{eqn:badprop}
	\Sigma = \begin{bmatrix}
		0.75^2 & 0 & 0 \\
		0 & 0.75^2 & 0 \\
		0 & 0 & 0.03^2
	\end{bmatrix}.
\end{eqnarray}
Such a proposal kernel is reasonable to characterise the initial explorations of an ABC posterior as no correlations between parameters are assumed.

While MLMC-ABC does not require a proposal kernel function, some knowledge of the variance of each bias correction term is needed to determine the optimal sample numbers, $N_1,\ldots,N_L$. This is achieved using $100$ trial samples of each level.  The number of data generation steps is also recorded to compute $N_1,\dots, N_L$, as per \eqref{eqn:optN}. The resulting sample numbers are then scaled such that $N_L$ is a user prescribed value.

The efficiency metric we use is the root mean-squared error (RMSE) of the CDF estimate versus the total number of data generation steps, $N_s$. The mean-squared error (MSE) is taken under the $L_\infty$ norm, that is, $\text{MSE} = \E{\|F_{\epsilon_\ell} - \hat{F}_{\epsilon_\ell}\|_\infty^2}$ where $F_{\epsilon_\ell}$ is the exact ABC posterior CDF and $\hat{F}_{\epsilon_\ell}$ is the Monte Carlo estimate using a regular lattice with $100 \times  100 \times 100$ points. To compute the RMSE, a high precision solution is computed using $10^6$ ABC rejection samples of the target ABC posterior. This is computed over a period of 48 hours using 500 processor cores.

\begin{figure*}
	\centering	
	\includegraphics[width=\textwidth]{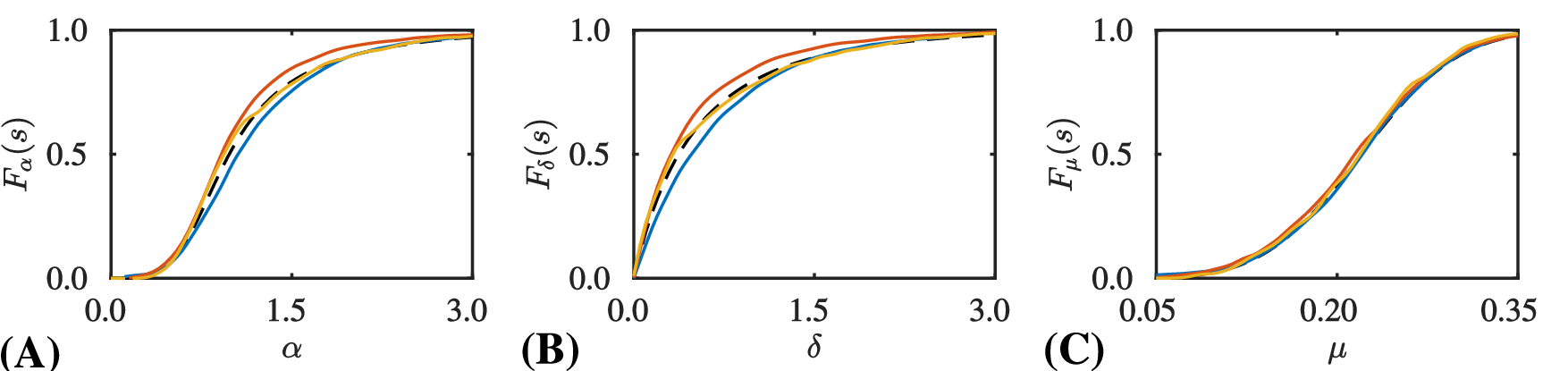}
	\caption{Estimated marginal CDFs--({\bf A}) $\alpha$, ({\bf B}) $\delta$ and ({\bf C}) $\mu$--for the tuberculosis transmission stochastic model. Estimate computed using using MLMC-ABC with $N_L = 800$ (solid yellow), MCMC-ABC over $1.2\times10^6$ iterations (solid blue), SMC-ABC with $3,200$ particles (solid red) and high precision solution (dashed black).}
	\label{fig:tanmarg}
\end{figure*}

Table~\ref{tab:notuneRMSE} presents the RMSE for MCMC-ABC, SMC-ABC and MLMC-ABC for different configurations. The RMSE values are computed using 10 independent estimator calculations. The algorithm parameter varied is the number of particles, $N_P$, for SMC-ABC, the number of iterations, $N_T$, for MCMC-ABC and the level $L$ sample number, $N_L$, for MLMC-ABC. 
\begin{table*}	
	\begin{tabular}{ccccccccc}
		\hline
		\multicolumn{3}{c}{MLMC-ABC} & \multicolumn{3}{c}{MCMC-ABC}  & \multicolumn{3}{c}{SMC-ABC}\\ %\multicolumn{2}{c}{MCMC-ABC}
		$N_L$ & $N_s$ &  RMSE & $N_T$ & $N_s$ &  RMSE & $N_P$ & $N_s$ &  RMSE \\
		\hline
		\hline
		$100$& $102,246$ & $0.1362$ &  $160,000$ & $163,800$ & $0.2434$ &$400$& $347,281$ & $0.1071$\\
		$200$& $255,443$ & $0.1136$ & $320,000$ & $323,841$ & $0.1828$ &$800$& $701,100$ & $0.0954$ \\
		$400$& $577,312$ & $0.1067$ & $640,000$ & $643,930$ & $0.1832$ &$1,600$& $1,385,790$ & $0.0858$ \\
		$800$& $1,040,140$ & $0.0861$ & $1,280,000$ & $1,283,590$ & $0.1345$ & $3,200$& $2,792,510$ & $0.0784$\\
		\hline
	\end{tabular}
	\centering
	\caption{Comparison of MLMC-ABC against MCMC-ABC and SMC-ABC using a naive proposal kernel.}
	\label{tab:notuneRMSE}
\end{table*}
Using the proposal kernel provided in \eqref{eqn:badprop}, SMC-ABC requires almost $30\%$ more data generation steps than MLMC-ABC to obtain the same RMSE. MLMC-ABC obtains nearly double the accuracy of MCMC-ABC for the same number of data generation steps.
Figure~\ref{fig:tanmarg} shows an example of the high precision marginal posterior CDFs, $F_{\alpha}(s)$, $F_{\delta}(s)$ and $F_\mu(s)$, compared with the numerical solutions computed using the three methods.

We note that these results represent a typical scenario  when solving this problem with a standard choice of proposal densities for MCMC-ABC and SMC-ABC. However, obtaining a good proposal kernel is a difficult open problem, and infeasible to do heuristically for high-dimensional parameter spaces. Therefore, efficient proposal kernels are almost never obtained without significant manual adjustment or additional algorithmic modifications such as adaptive schemes \citep{Beaumont2009,DelMoral2012,Roberts2009}. Nevertheless, we demonstrate in Table~\ref{tab:tuneRMSE} the increased efficiency for MCMC-ABC and SMC-ABC when using a highly configured Gaussian proposal kernel with covariance matrix
as determined by \cite{Tanaka2006}
\begin{eqnarray}
	\label{eqn:prop}
	\Sigma = \begin{bmatrix}
		0.5^2 & 0.225 & 0 \\
		0.225 & 0.5^2 & 0 \\
		0 & 0 & 0.015^2
	\end{bmatrix}.
\end{eqnarray}

\begin{table*}
	\begin{tabular}{ccccccccc}
		\hline
		\multicolumn{3}{c}{MLMC-ABC} & \multicolumn{3}{c}{MCMC-ABC}  & \multicolumn{3}{c}{SMC-ABC}\\ %\multicolumn{2}{c}{MCMC-ABC}
		$N_L$ & $N_s$ &  RMSE & $N_T$ & $N_s$ &  RMSE & $N_P$ & $N_s$ &  RMSE \\
		\hline
		\hline
		$100$& $102,246$ & $0.1362$ & $160,000$ & $161,604$ & $0.1718$ &$400$& $102,902$ & $0.1270$\\
		$200$& $255,443$ & $0.1136$ & $320,000$ & $322,962$ & $0.1254$ &$800$& $198,803$ & $0.0767$ \\
		$400$& $577,312$ & $0.1067$ & $640,000$ & $642,412$ & $0.1127$   &$1,600$& $402,334$ & $0.0652$ \\
		$800$& $1,040,140$ & $0.0861$ & $1,280,000$    &  $1,305,340$  & $0.0934$ & $3,200$& $797,893$ & $0.0560$\\
		
		\hline
	\end{tabular}
	\centering
	\caption{Comparison of MLMC-ABC against MCMC-ABC and SMC-ABC using heuristically chosen proposal densities.}
	\label{tab:tuneRMSE}
\end{table*}
We note that \cite{Sisson2007} use a similar proposal kernel, however they do not explicitly state the difference in the covariance matrix $\Sigma$; we therefore assume that \eqref{eqn:prop} represents a proposal kernel that is heuristically optimal to the target posterior density. We emphasise that it would be incredibly rare to arrive at an optimal proposal kernel without any additional experimentation.
Even in this unlikely case, MLMC-ABC is still comparable with MCMC-ABC. However, MLMC-ABC is clearly not as efficient as SMC-ABC for this heuristically optimised scenario. The scenario is intentionally biased toward MCMC-ABC and SMC-ABC, so this result is not unexpected. Future research could consider a comparison of the methods when the extra computational burden of determining the optimal $\Sigma$, such as implementation of an adaptive scheme, is taken into account.

\section{Discussion}
\label{sec:discuss}
Our results indicate that, while SMC-ABC and MCMC-ABC can be heuristically optimised to be highly efficient, an accurate estimate of the parameter posterior can be obtained using the MLMC-ABC method presented here in a relatively automatic fashion. Furthermore, the efficiency of MLMC-ABC is comparable or improved over MCMC-ABC and SMC-ABC, even in the case when efficient proposal kernels are employed.

The need to estimate the variances of each bias correction term could be considered a limitation of the MLMC-ABC approach. However, we find in practice that these need not be computed to high accuracy and can often be estimated with a relatively small number of samples. There could be examples of Bayesian inference problems where MLMC-ABC is inefficient on account of the variance estimation inaccuracy. We have so far, however, failed to find an example for which $100$ samples of each bias correction term is insufficient to obtain a good MLMC-ABC estimator.

There are many modifications one could consider to further improve MLMC-ABC. In this work, we explicitly specify the sequence of acceptance thresholds in advance for both MLMC-ABC and SMC-ABC. As this is the initial presentation of the method, it is appropriate to consider this idealised case. However, it is unclear if an optimal sequence of thresholds for SMC-ABC will also be optimal for MLMC-ABC and vice versa.  Furthermore, as mention in Section~\ref{sec:intro}, practical applications of SMC-ABC often determine such sequences adaptively~\citep{Drovandi2011,Silk2013}. A modification of MLMC-ABC to allow for adaptive acceptance thresholds would make MLMC-ABC even more practical as it could be used to minimise the coupling bias. The exact mechanisms to achieve this could be based on similar ideas to adaptive SMC-ABC~\citep{Drovandi2011}. Given the solution at level $\ell-1$, the next level $\ell$ could be determined through: (i) sampling a fixed set of \emph{prior} samples; (ii) sorting these samples based on the discrepancy metric; (iii) selecting a discrepancy threshold, $\epsilon_\ell$, that optimises coupling bias and the variance of the bias correction term. Future work should address these open problems.

Other improvements could focus on the discretisation used for the eCDF calculations. The MLMC-ABC method has no requirement of a regular lattice, and alternative choices would likely enable MLMC-ABC to scale to much higher dimensional parameter spaces. Adaptive grids that refine with each level is one option that could be considered; however, unstructured grids or kernel based methods also have potential.

Improvements to the coupling scheme are also possible avenues for future consideration. The coupling approach we have considered depends only on the computation of the marginal posterior CDFs and assumes nothing about the underlying model. It may be possible to take advantage of certain model specific features to improve variance reduction. There may also be cases where the rejection sampling scheme is prohibitive for the more accurate acceptance levels. The combination of our coupling scheme with the MLSMC scheme recently proposed by~\cite{Beskos2016} and~\cite{Jasra2017} is a promising possibility to mitigate these issues. Future work should investigate, compare and contrast the variety of available coupling strategies in the growing body of literature on MLMC for ABC and Bayesian inverse problems.

We have shown, in a practical way, how MLMC techniques can be applied to ABC inference. We also demonstrate that variance reduction strategies, even when applied to simple methods such as rejection sampling, can achieve performance improvements comparable, and in some cases superior, to modern advanced ABC methods based on MCMC and SMC methods. Therefore, the MLMC framework is a promising area for designing improved samplers for complex statistical problems with intractable likelihoods.

\paragraph{{\bf Acknowledgements}}
This work is supported by the Australian Research Council (FT130100148). REB would like to thank the Royal Society for a Royal Society Wolfson Research Merit Award and the Leverhulme Trust for a Leverhulme Research Fellowship. REB would also like to acknowledge support from the Biotechnology and Biological Sciences Research Council (BB/R000816/1). Computational resources were provided by High Performance Computing and Advanced Research Computing support (HPC-ARCs), Queensland University of Technology (QUT). The authors thank Mike Giles and Chris Drovandi for helpful discussions, and two anonymous reviewers and the Associate Editor for insightful comments and suggestions.

\begin{appendices}
\appendix

\section{MLMC for ABC with general Lipschitz functions}
\label{sec:appdx}
Minor modifications of the MLMC-ABC method (Algorithm~\ref{alg:mlmc-abc}) are possible to enable the computation of expectations of the form
\begin{equation*}
	\Econd{U(\paramvec)}{\discrep{\dat}{\simdat} < \epsilon} = \int_{\paramspace} U(\paramvec) \CondPDF{\paramvec}{\discrep{\dat}{\simdat} < \epsilon} \, \text{d}\param_k \ldots \text{d}\param_1,
\end{equation*}
where $U(\paramvec)$ is any Lipschitz continuous function.

Using the same notation as defined in Section~\ref{sec:mlmc-est-form}, the MLMC telescoping sum may be formed for a given sequence of acceptance thresholds, $\epsilon_1 > \cdots > \epsilon_L$, to compute the expectation $E_{\epsilon_L} = \E{U(\paramvec_{\epsilon_L})}$. That is,
\begin{equation}
	\label{alg:mlmc-abc-tel-gen}
	E_{\epsilon_L} = \sum_{\ell=1}^{L} P_{\epsilon_\ell}, \quad P_{\epsilon_\ell} = \begin{cases}
		\E{U(\paramvec_{\epsilon_1})}, & \ell = 1, \\
		\E{U(\paramvec_{\epsilon_\ell}) - U(\paramvec_{\epsilon_{\ell-1}})}, & \ell > 1.
	\end{cases}
\end{equation}
From \eqref{alg:mlmc-abc-tel-gen}, it is straightforward to obtain equivalent expressions to \eqref{eqn:mlmcEst} and \eqref{eqn:bMCest}.
\begin{algorithm}
	\caption{Modified MLMC-ABC}
	\begin{algorithmic}
		\State Initialise $\epsilon_1,\ldots, \epsilon_L$, $N_1,\dots, N_L$ and prior $\PDF{\paramvec}$.
		\State Set $\PDF{\paramvec_{\epsilon_0}} \leftarrow \PDF{\paramvec}$.
		\For{$\ell = 1,\ldots, L$}
		\For{$i = 1, \ldots, N_\ell$}
		\Repeat
		\State Sample $\paramvec^{*} \sim \PDF{\paramvec}$ restricted to $\supp(\PDF{\paramvec_{\epsilon_{\ell-1}}})$.		
		\State Generate data, $\simdat \sim \simProc{\dat}{\paramvec^*}$.
		\Until{$\discrep{\dat}{\simdat} \leq \epsilon_\ell$.}
		\State Set $\paramvec_{\epsilon_\ell}^{i} \leftarrow \paramvec^{*}$.
		\EndFor
		\For{$j = 1,\ldots,k$}
		\For{$s \in \mathcal{S}_j$}
		\State Set $\hat{F}_{\epsilon_\ell,j}^{N_\ell}(s) \leftarrow \sum_{i=1}^{N_\ell} \xi\left((s_j - \param_{\epsilon_\ell,j}^{i})/\delta_j\right)/N_\ell$.
		\EndFor
		\EndFor
		\If{$\ell = 1$}
		\State $\hat{E}_{\epsilon_\ell}^{N_\ell} \leftarrow \sum_{i=1}^{N_\ell} U(\paramvec^{i}_{\epsilon_\ell})/N_\ell$
		\Else
		\For{$i = 1, \ldots, N_\ell$}
		
		\For{$j = 1,\ldots, k$}
		\State Set $\param_{\epsilon_{\ell-1},j}^{i} \leftarrow \hat{G}^{N_1,\dots,N_{\ell-1}}_{\epsilon_{\ell-1},j}\left(\hat{F}_{\epsilon_\ell,j}^{N_\ell}\left(\param_{\epsilon_\ell,j}^{i}\right)\right)$.
		\EndFor
		\EndFor
		\For{$j = 1, \ldots, k$}
		\For{$s \in \mathcal{S}_j$}
		\State Set $\hat{Y}^{N_\ell}_{\epsilon_\ell,j}(s) \leftarrow \sum_{i=1}^{N_\ell} \left[\xi\left((s_j - \param_{\epsilon_\ell,j}^{i})/\delta_j\right) - \xi\left((s_j - \param_{\epsilon_{\ell-1},j}^{i})/\delta_j\right)\right]/N_\ell$.
		\State Set $\hat{F}^{N_1,\ldots,N_\ell}_{\epsilon_\ell,j}(s) \leftarrow \hat{F}^{N_1,\ldots,N_{\ell-1}}_{\epsilon_{\ell-1},j}(s)+ \hat{Y}^{N_\ell}_{\epsilon_\ell,j}(s)$.
		\EndFor
		\EndFor
		\State $\hat{P}_{\epsilon_\ell}^{N_\ell} \leftarrow \sum_{i=1}^{N_\ell} \left[U(\paramvec^{i}_{\epsilon_\ell}) - U(\paramvec^{i}_{\epsilon_{\ell-1}}))\right]/N_\ell$
		\State $\hat{E}^{N_1,\ldots,N_\ell}_{\epsilon_\ell} \leftarrow \hat{E}^{N_{1},\ldots,N_{\ell-1}}_{\epsilon_{\ell-1}} + \hat{P}_{\epsilon_\ell}^{N_\ell}$.
		\EndIf
		\EndFor
	\end{algorithmic}
	\label{alg:mlmc-abc-gen}
\end{algorithm}

The modified MLMC-ABC algorithm proceeds in a very similar fashion to Algorithm~\ref{alg:mlmc-abc}, however, there is no need to hold a complete discretisation of the parameter space, $\paramspace$. This is because only the $k$ marginal CDFs, $F_{\epsilon_\ell,1}\left(s_1\right), \ldots,F_{\epsilon_\ell,k}\left(s_k\right)$, are required to form the coupling strategy in Section~\ref{sec:VarReduc}. Thus, we denote $\mathcal{S}_j$ to be a discretisation of the $j$th coordinate axis. This significantly reduces the computational burden of the lattice in higher dimensions since $|\mathcal{S}_j| = \mathcal{O}(\log_k |\mathcal{S}|)$. The resulting algorithm is given in Algorithm~\ref{alg:mlmc-abc-gen}.
\end{appendices}


\begin{thebibliography}{36}
	\providecommand{\natexlab}[1]{#1}
	\providecommand{\url}[1]{{#1}}
	\providecommand{\urlprefix}{URL }
	\expandafter\ifx\csname urlstyle\endcsname\relax
	\providecommand{\doi}[1]{DOI~\discretionary{}{}{}#1}\else
	\providecommand{\doi}{DOI~\discretionary{}{}{}\begingroup
		\urlstyle{rm}\Url}\fi
	\providecommand{\eprint}[2][]{\url{#2}}
	
	\bibitem[{Anderson and Higham(2012)}]{Anderson2012}
	Anderson, D.F., Higham, D.J., 2012. Multilevel Monte Carlo for continuous time Markov
	chains, with applications in biochemical kinetics. Multiscale Modeling \&
	Simulation 10(1):146--179. \doi{10.1137/110840546}
	
	
	\bibitem[Avikainen(2009)]{Avikainen2009}
	Avikainen, R., 2009. On irregular functionals of SDEs and the Euler scheme. Finance and Stochastics 13(3):381--401.
	\doi{10.1007/s00780-009-0099-7}
	
	
	\bibitem[{Barber et~al.(2015)Barber, Voss, and Webster}]{Barber2015}
	Barber, S., Voss, J., Webster, M., 2015. The rate of convergence for approximate
	Bayesian computation. Electronic Journal of Statistics 9:80--105.
	\doi{10.1214/15-EJS988}
	
	\bibitem[{Beaumont et~al.(2002)Beaumont, Zhang, and Balding}]{Beaumont2002}
	Beaumont, M.A., Zhang, W., Balding, D.J., 2002. Approximate Bayesian computation in
	population genetics. Genetics 162(4):2025--2035.
	
	\bibitem[{Beaumont et~al.(2009)Beaumont, Cornuet, Marin, and
		Robert}]{Beaumont2009}
	Beaumont, M.A., Cornuet, J.M., Marin, J.M., Robert, C.P., 2009. Adaptive approximate
	Bayesian computation. Biometrika 96(4):983--990.
	
	\bibitem[{Beskos et~al.(2017)Beskos, Jasra, Law, Tempone, and Zhou}]{Beskos2016}
	Beskos, A., Jasra, A., Law, K., Tempone, R., Zhou, Y., 2017. Multilevel sequential Monte
	Carlo samplers. Stochastic Processes and their Applications 127(5):1417--1440.
	\doi{10.1016/j.spa.2016.08.004}
	
	\bibitem[{Browning et~al.(2018)Browning, McCue, Binny, Planck, Shah, and Simpson}]{Browning2017}
	Browning, A.P., McCue, S.W., Binny, R.N., Plank, M.J., Shah, E.T., Simpson, M.J., 2018. Inferring parameters for a lattice-free model of cell migration and proliferation using experimental data. Journal of Theoretical Biology 437:251--260.
	\doi{10.1016/j.jtbi.2017.10.032}

	
	\bibitem[Cabras et~al.(2015)Cabras, Castellanos Neuda, and Ruli]{Cabras2015}
	Cabras, S., Castellanos Neuda, M.E., Ruli, E., 2015. Approximate Bayesian computation by modelling summary statistics in a quasi-likelihood framework. Bayesian Analysis 10(2):411--439.
	\doi{10.1214/14-BA921}
	
	
	\bibitem[{Del~Moral et~al.(2006)Del~Moral, Doucet, and Jasra}]{DelMoral2006}
	Del~Moral, P., Doucet, A., Jasra, A., 2006. Sequential Monte Carlo samplers. Journal
	of the Royal Statistical Society: Series B (Statistical Methodology)
	68(3):411--436. \doi{10.1111/j.1467-9868.2006.00553.x}
	
	\bibitem[{Del~Moral et~al.(2012)Del~Moral, Doucet, and Jasra}]{DelMoral2012}
	Del~Moral, P., Doucet, A., Jasra, A., 2012. An adaptive sequential Monte Carlo method
	for approximate Bayesian computation. Statistics and Computing
	22(5):1009--1020. \doi{10.1007/s11222-011-9271-y}
	
	\bibitem[{Dodwell et~al.(2015)Dodwell, Ketelsen, Scheichl, and
		Teckentrup}]{Dodwall2015}
	Dodwell, T.J., Ketelsen, C., Scheichl, R., Teckentrup, A.L., 2015. A hierarchical
	multilevel Markov chain Monte Carlo algorithm with applications to
	uncertainty quantification in subsurface flow. SIAM/ASA Journal on
	Uncertainty Quantification 3(1):1075--1108. \doi{10.1137/130915005}
	
	\bibitem[{Drovandi and Pettitt(2011)}]{Drovandi2011}
	Drovandi, C.C., Pettitt, A.N., 2011. Estimation of parameters for macroparasite
	population evolution using approximate Bayesian computation. Biometrics
	67(1):225--233. \doi{10.1111/j.1541-0420.2010.01410.x}
	
	\bibitem[{Efendiev et~al.(2015)Efendiev, Jin, Michael, and Tan}]{Efendiev2015}
	Efendiev, Y., Jin, B., Michael, P., Tan, X., 2015. Multilevel Markov chain Monte Carlo
	method for high-contrast single-phase flow problems. Communications in
	Computational Physics 17(1):259--286. \doi{10.4208/cicp.021013.260614a}
	
	\bibitem[{Fearnhead and Prangle(2012)}]{Fearnhead2012}
	Fearnhead, P., Prangle, D., 2012. Constructing summary statistics for approximate
	Bayesian computation: semi-automatic approximate Bayesian computation.
	Journal of the Royal Statistical Society Series B (Statistical Methodology)
	74(3):419--474. \doi{10.1111/j.1467-9868.2011.01010.x}
	
	\bibitem[{Filippi et~al.(2013)Filippi, Barnes, Cornebise, and
		Stumpf}]{Filippi2013}
	Filippi, S., Barnes, C.P., Cornebise, J., Stumpf, M.P.H., 2013. On optimality of kernels
	for approximate Bayesian computation using sequential Monte Carlo.
	Statistical Applications in Genetics and Molecular Biology 12(1):87--107.
	\doi{10.1515/sagmb-2012-0069}
	
	\bibitem[{Giles(2008)}]{Giles2008}
	Giles, M.B., 2008. Multilevel Monte Carlo path simulation. Operations Research
	56(3):607--617. \doi{10.1287/opre.1070.0496}
	
	\bibitem[{Giles et~al.(2015)Giles, Nagapetyan, and Ritter}]{Giles2015b}
	Giles, M.B., Nagapetyan, T., Ritter, K., 2015. Multilevel Monte Carlo approximation of
	cumulative distribution function and probability densities. SIAM/ASA Journal
	on Uncertainty Quantification 3(1):267--295. \doi{10.1137/140960086}
	
	\bibitem[{Gillespie(1977)}]{Gillespie1977}
	Gillespie, D.T., 1977. Exact stochastic simulation of coupled chemical reactions.
	The Journal of Physical Chemistry 81(25):2340--2361.
	\doi{10.1021/j100540a008}
	
	\bibitem[{Green et~al.(2015)Green, {\L}atuszy{\'{n}}ski, Pereyra, and
		Robert}]{Green2015}
	Green, P.J., {\L}atuszy{\'{n}}ski, K., Pereyra, M., Robert, C.P., 2015. Bayesian
	computation: a summary of the current state, and samples backwards and
	forwards. Statistics and Computing 25(4):835--862.
	\doi{10.1007/s11222-015-9574-5}
	
	\bibitem[{Gregory et~al.(2016)Gregory, Cotter, and Reich}]{Gregory2016}
	Gregory, A., Cotter, C.J., Reich, S., 2016. Multilevel ensemble transform particle
	filtering. SIAM Journal on Scientific Computing 38(3):A1317--A1338.
	\doi{10.1137/15M1038232}
	
	
	\bibitem[Grelaud et~al.(2009)Grelaud, Marin, Robert, Rodolphe, and Tally]{Grelaud2009}
	Grelaud, A., Marin, J.-M., Robert, C.P., Rodolphe, F., Tallay, J.-F., 2009. ABC likelihood-free methods for model choice in Gibbs random fields. Bayesian Analysis 4(2):317--335.
	\doi{10.1214/09-BA412}
	
	
	
	\bibitem[{Guha and Tan(2017)}]{Guha2017}
	Guha, N., Tan, X., 2017. Multilevel approximate Bayesian approaches for flows in
	highly heterogeneous porous media and their applications. Journal of
	Computational and Applied Mathematics 317:700 -- 717.
	\doi{10.1016/j.cam.2016.10.008}
	
	\bibitem[{Hastings(1970)}]{Hastings1970}
	Hastings, W.K., 1970. Monte Carlo sampling methods using Markov chains and their
	applications. Biometrika 57(1):97--109.
	
	\bibitem[{Jasra et~al.(2017)Jasra, Jo, Nott, Shoemaker, and Tempone}]{Jasra2017}
	Jasra, A., Jo, S., Nott, D., Shoemaker, C., Tempone, R., 2017. Multilevel Monte Carlo in
	approximate Bayesian computation. ArXiv e-prints
	\urlprefix\url{https://arxiv.org/abs/1702.03628}, \eprint{1702.03628}.
	

	\bibitem[Johnston et~al.(2014)Johnston, Simpson, McElwain, Binder, and Ross]{Johnston2014}
	Johnston, S.T., Simpson, M.J., McElwain, D.L.S., Binder, B.J., Ross, J.V., 2014 Interpreting scratch assays using pair density dynamics and approximate Bayesian computation. Open Biology 4(9):140097.
	\doi{10.1098/rsob.140097}

	
	\bibitem[{Lester et~al.(2016)Lester, Baker, Giles, and Yates}]{Lester2014}
	Lester, C., Baker, R.E., Giles, M.B., Yates, C.A., 2016. Extending the multi-level method
	for the simulation of stochastic biological systems. Bulletin of Mathematical
	Biology 78(8):1640--1677. \doi{10.1007/s11538-016-0178-9}
	
	\bibitem[{Marjoram et~al.(2003)Marjoram, Molitor, Plagnol, and
		Tavar{\'e}}]{Marjoram2003}
	Marjoram, P., Molitor, J., Plagnol, V., Tavar{\'e}, S., 2003. Markov chain Monte Carlo
	without likelihoods. Proceedings of the National Academy of Sciences of the
	United States of America 100(26):15,324--15,328.
	\doi{10.1073/pnas.0306899100}
	
	
	\bibitem[{Marin et~al.(2012)Marin, Pudlo, Robert, and Ryder}]{Marin2012}
	Marin, J.-M., Pudlo, P., Robert, C.P., Ryder, R.J., 2012. Approximate Bayesian computational methods. Statistics and Computing 22(6):1167--1180.
	\doi{10.1007/s11222-011-9288-2}
	
	
	\bibitem[{Metropolis et~al.(1953)Metropolis, Rosenbluth, Rosenbluth, Teller, and
		Teller}]{Metroplis1953}
	Metropolis, N., Rosenbluth, A.W., Rosenbluth, M.N., Teller, A.H., Teller, E., 1953. Equation
	of state calculations by fast computing machines. The Journal of Chemical
	Physics 21(6):1087--1092. \doi{10.1063/1.1699114}
	
	
	\bibitem[{Navascu\'{e}s et~al.(2017)Navascu\'{e}s, Leblois, and Burgarella}]{Navascues2017}
	Navascu\'{e}s, M., Leblois, R., Burgarella, C., 2017. Demographic inference through approximate-Bayesian-computation skyline plots. PeerJ 5:e3530.
	\doi{10.7717/peerj.3530}
	
	
	\bibitem[{Pooley et~al.(2015)Pooley, Bishop, and Marion}]{Pooley2015}
	Pooley, C.M., Bishop, S.C., Marion, G., 2015. Using model-based proposals for fast
	parameter inference on discrete state space, continuous-time Markov
	processes. Journal of the Royal Society Interface 12(107):20150225.
	\doi{10.1098/rsif.2015.0225}
	
	
	\bibitem[{Pritchard et~al.(1999)}]{Pritchard1999}
	Pritchard, J.K, Seielstad, M.T., Perez-Lezaun, A.. Feldman, M.W., 1999. Population growth of human Y chromosomes: a study of Y chromosome microsatellites.
	Molecular Biology and Evolution
	16(12):1791--1798.
	\doi{10.1093/oxfordjournals.molbev.a026091}
	
	
	\bibitem[{Reiss(1981)Reiss}]{Reiss1981}
	Reiss, R.-D., 1981. Nonparametric estimation of smooth distribution functions. Scandinavian Journal of Statistics 8:116--119. 
	
	
	
	
	\bibitem[{Roberts and Rosenthal(2004)}]{Roberts2004}
	Roberts, G.O., Rosenthal, J.S., 2004. General state space Markov chains and MCMC
	algorithms. Probability Surveys 1:20--71. \doi{10.1214/154957804100000024}
	
	
	\bibitem[{Roberts and Rosenthal(2009)}]{Roberts2009}
	Roberts, G.O., Rosenthal, J.S., 2009. Examples of adaptive MCMC. Journal of
	Computational and Graphical Statistics 18(2):349--367.
	\doi{10.1198/jcgs.2009.06134}
	
	\bibitem[{Ross et~al.(2017)Ross, Baker, Parker, Ford, Mort, and Yates}]{Ross2017}
	Ross, R.J.H., Baker, R.E., Parker, A., Ford, M.J., Mort, R.L., Yates, C.A., 2017.
	Using approximate Bayesian computation to quantify cell-cell adhesion parameters
	in cell migratory process. npj Systems Biology and Applications 3(1):9.
	\doi{10.1038/s41540-017-0010-7}
	
	
	\bibitem[Silk et~al.(2013)Silk, Filippi, and Stumpf]{Silk2013}
	Silk, D., Filippi, S., Stumpf, M.P.H., 2013. Optimizing threshold-schedules for sequential approximate Bayesian computation: applications to molecular systems. Statistical Applications in Genetics and Molecular Biology 12(3):603--618.
	\doi{10.1515/sagmb-2012-004}
	
	
	
	\bibitem[Silverman(1986)Silverman]{Silverman1986}
	Silverman, B.W., 1986. Density estimation for statistics and data analysis. Monographs on Statistics and Applied Probability. Boca Ranton: Chapman \& Hall/CRC.
	
	
	\bibitem[{Sisson et~al.(2007)Sisson, Fan, and Tanaka}]{Sisson2007}
	Sisson, S.A., Fan, Y., Tanaka, M.M., 2007. Sequential Monte Carlo without likelihoods.
	Proceedings of the National Academy of Sciences of the United States of
	America 104(6):1760--1765. \doi{10.1073/pnas.0607208104}
	
	\bibitem[{Small et~al.(1994)Small, Hopewell, Singh, Paz, Parsonnet, Ruston,
		Schecter, Daley, and Schoolnik}]{Small1994}
	Small, P.M., Hopewell, P.C., Singh, S.P., Paz, A., Parsonnet, J., Ruston, D.C., Schecter, G.F.,
	Daley, C.L., Schoolnik, G.K., 1994. The epidemiology of tuberculosis in San
	Francisco -- a population-based study using conventional and molecular
	methods. New England Journal of Medicine 330(24):1703--1709.
	\doi{10.1056/NEJM199406163302402}
	
	\bibitem[{Stumpf(2014)}]{Stumpf2014}
	Stumpf, M.P.H., 2014. Approximate Bayesian inference for complex ecosystems.
	F1000Prime Reports 6:60. \doi{10.12703/P6-60}
	
	\bibitem[{Sunn{\aa}ker et~al.(2013)Sunn{\aa}ker, Busetto, Numminen, Corander, Foll, and
		Dessimoz}]{Sunnaker2013}
	Sunn{\aa}ker, M., Busetto, A.G., Numminen, E., Corander, J., Foll, M., Dessimoz, C., 2013.
	Approximate Bayesian computation. PLOS Computational Biology 9(1):e1002803.
	\doi{10.1371/journal.pcbi.1002803}
	
	\bibitem[{Tanaka et~al.(2006)Tanaka, Francis, Luciani, and Sisson}]{Tanaka2006}
	Tanaka, M.M., Francis, A.R., Luciani, F., Sisson, S.A., 2006. Using approximate Bayesian
	computation to estimate tuberculosis transmission parameter from genotype
	data. Genetics 173(3):1511--1520. \doi{10.1534/genetics.106.055574}
	
	\bibitem[{Tavar{\'e} et~al.(1997)Tavar{\'e}, Balding, Griffiths, and
		Donnelly}]{Tavare1997}
	Tavar{\'e}, S., Balding, D.J., Griffiths, R.C., Donnelly, P., 1997. Inferring coalescence
	times from DNA sequence data. Genetics 145(2):505--518.
	
	\bibitem[{Thorne and Stumpf(2012)}]{Thorne2012}
	Thorne, T., Stumpf, M.P.H., 2012. Graph spectral analysis of protein interaction
	network evolution. Journal of The Royal Society Interface 9(75):2653--2666.
	\doi{10.1098/rsif.2012.0220}
	
	\bibitem[{Toni et~al.(2009)Toni, Welch, Strelkowa, Ipsen, and Stumpf}]{Toni2009}
	Toni, T., Welch, D., Strelkowa, N., Ipsen, A., Stumpf, M.P.H., 2009. Approximate Bayesian
	computation scheme for parameter inference and model selection in dynamical
	systems. Journal of the Royal Society Interface 6:187--202.
	\doi{10.1098/rsif.2008.0172}
	
	\bibitem[{Vo et~al.(2015)Vo, Drovandi, Pettit, and Simpson}]{Vo2015}
	Vo, B.N., Drovandi, C.C., Pettit, A.N., Simpson, M.J., 2015. Quantifying uncertainty in
	parameter estimates for stochastic models of collective cell spreading using
	approximate Bayesian computation. Mathematical Biosciences 263:133--142.
	\doi{10.1016/j.mbs.2015.02.010}
	
	\bibitem[Warne et~al.(2017)]{Warne2017}
	Warne, D.J., Baker, R.E., Simpson M.J., 2017. Optimal quantification of contact inhibition in cell populations. Biophysical Journal 113(9):1920--1924.
	\doi{10.1016/j.bpj.2017.09.016}
	
	\bibitem[{Weiss and Dishon(1971)}]{Weiss1971}
	Weiss, G.H., Dishon, M., 1971. On the asymptotic behavior of the stochastic and deterministic models of an epidemic.
	Mathematical Biosciences 11(3):261--265.
	\doi{10.1016/0025-5564(71)90087-3}
	
	\bibitem[{Wilkinson(2009)}]{Wilkinson2009}
	Wilkinson, D.J., 2009. Stochastic modelling for quantitative description of
	heterogeneous biological systems. Nature Reviews Genetics 10:122--133.
	\doi{10.1038/nrg2509}
	
\end{thebibliography}

%\section*{References}

\end{document}